\documentstyle[prl,aps,amssymb,amsfonts,twocolumn,epsf]{revtex}
\def\1#1{{\bf #1}}
\def\2#1{{\cal #1}}
\def\4#1{{\tt #1}}
\def\5#1{{\sf #1}}
\def\6#1{{\frak #1}}
\def\7#1{{\Bbb #1}}
\def\8#1{{\rm #1}}
\def\9#1{{\cal #1}}
\newtheorem{The}{Theorem}[section]

\newtheorem{Pro}[The]{Proposition}

\newtheorem{Alg}[The]{Algorithm}
  
\title{\bf Logical network implementation for cluster states and graph codes}
\author{Dirk Schlingemann \\
 {\small Institut f{\"u}r Mathematische Physik, TU Braunschweig,}\\
  {\small Mendelssohnstr.3, 38106 Braunschweig, Germany.}}
\begin{document}
\draft \maketitle
\narrowtext
\abstract{In a previous paper a straight forward construction
method for quantum error correcting codes,
based on graphs, has been presented.
These graph codes are directly related to cluster states
which have been introduced by Briegel and Raussendorf.
We show that the preparation of a
cluster state as well as the coding operation for a graph code,
can be implemented by a logical
network. Concerning the qubit case
each vertex corresponds to an Hadamard gate and each edge
corresponds to a controlled not gate.}

\section{Introduction}
Recently, we discussed a new construction method for quantum
error correcting codes, based on graphs \cite{SchlWer00}.
The error correcting capabilities of a "graph code" can directly
be derived from the structure of the graph which also allows to
tailor quantum error correcting codes systematically.

In the present paper we discuss how graph codes can be
implemented on a quantum computer by a logical network.
This problem has also been analyzed by several authors with
respect to quantum error correcting codes being obtained
by other construction methods \cite{ClGot96,Grassl}.

The subsequent results are based on a direct relation between
graph codes and the so called "cluster states".
These states have been introduced by H.J. Briegel and R.
Raussendorf for performing quantum algorithms by means of local
von Neumann measurements \cite{BrieRau00,BrieRau01}.
Besides these aspects, cluster states can also be used for studying
concepts multi-particle entanglement \cite{BrieRau01b}.

A cluster state is determined by the following objects:
\begin{itemize}
\item
A finite number of levels,
\item
A graph consisting of a finite number of vertices and edges, where
two vertices are connected by at most one edge
\item
Non-zero integral numbers (weights), each of them being
attached to an edge.
\end{itemize}
The number of levels is the dimension of a
single elementary quantum system, which we call here a "quantum digit".
For example,
considering an atom or ion in a trap, the states of a quantum digit
are density matrices on a Hilbert space spanned
by finitely many
energy eigenvectors, each of them belonging to an eigenvalue of
multiplicity one. A two level or binary quantum digit
is just a qubit.

A cluster state corresponds to a system of multiple quantum digits
whose positions are labeled by the vertices of the graph.
Each edge of the graph can be viewed as an interaction between
the quantum digits corresponding to vertices which are connected by it.
The non-zero integral number (weight) attached to the edge
can be viewed as the strength of the interaction.
As example, one may think of a two dimensional optical lattice structure,
where each quantum digit occupy a point within a two dimensional
cubic lattice with next neighbor interaction.

As far as operations on an optical lattice are concerned,
the preparation of
a cluster state needs only one elementary
operation applied to an initial product state -- each qubit is in a
superposition of "0" and "1".
This operation acts globally on the lattice, creating entanglement
between neighbored qubits.
Once a cluster state is prepared, it can be used as a resource for
performing any quantum algorithm by local measurements \cite{BrieRau00}.

Procedures for preparing a cluster state can also be given in
terms of one and two qubit gates. For any system of finitely many
qubits on which local Hadamard gates and CNOT gates
can be performed in a controlled manner,
we show in Section \ref{SI} that a cluster state
for an arbitrary graph
can be prepared by applying a logical network to the initial
state where each qubit is prepared in the state "0".
For a graph with $v$ vertices and $l$ edges,
the number of steps which are needed to perform the
corresponding network is $l+v$:
For each vertex there is a Hadamard gate and for each edge a CNOT
gate. This network can be derived from the graph by a systematic
algorithm.

This also opens the discussion for comparing
complexity measures for quantum algorithms
based on one- and two-qubit gates, on one hand, and
globally parallel operations, on the other hand.

Graph codes \cite{SchlWer00} are directly related to cluster states.
Consider the cluster state for a graph, we select for each
quantum digit, we wish to encode, a vertex. These selected
vertices are called "input vertices". The remaining vertices
are called "output vertices".
As we shall see in Section \ref{SII}, the coding operation for a graph
code can be split into four main steps. For the qubit case, it
works as follows:
\begin{enumerate}
\item[1:]
The input qubits are prepared in the state we wish to protect
against errors.
\item[2:]
Each output qubit is prepared in the state "0".
\item[3:]
A discrete dynamics is applied which creates entanglement between those qubits
sitting at vertices connected by an edge. Here it is given by first
performing a Hadamard transform on each input qubit, then applying
the network which creates the cluster state and finally one
operates again with a Hadamard transform on each input qubit.
\item[4:]
Each input qubit is measured in the "computational basis",
i.e. the basis given by the states "0" and "1".
\end{enumerate}

For a graph with $k$ input vertices, $n$ output vertices
and $l$ edges, as we already know,
the network for creating a cluster state can be realized by $k+n+l$
elementary gate operations. Implementing the dynamics of
step 3 would  cost at most $3k+n+l$ steps
since there are two additional Hadamard gates for each input.
Provided there are no edges between input vertices,
the number of gates can be reduced to $k+n+l$
as we shall see in Section \ref{SIII}. The
corresponding logical network can be
derived systematically from the graph analogously to the network
for cluster states.

Considering the case where only one input digit is present,
we derive logical networks for graph codes which
operate on the output digits. In comparison to the networks,
described before, it makes use of less resources.
We give in Section \ref{SIV} an algorithm associating to a graph with $n$
output vertices and $l$ edges a logical network which implements
the corresponding code with $n+l-1$ elementary gate operations.

\section{Logical networks for cluster states}
\label{SI}
To begin with, we briefly describe here how a
cluster state is constructed from
a finite number of levels $d$ and a weighted graph
$\Gamma$ with vertices $V$.
The classical configuration space for describing a digit is given
by a cyclic group $\7Z_d$. The states of a single quantum digit are
density operators
on the Hilbert space of "wave functions" $l_2(\7Z_d)$ \cite{l2} on $\7Z_d$.
Obviously, the quantum digit for the
two elementary group $d=2$ is just a qubit.

In our context a quantum register consists of
quantum digits labeled by
a finite set $V$ of "positions".
Thus the states of the quantum register
are density operators on
$l_2(\7Z_d^V)$ where $\7Z_d^V$ is the
group of tuples $\1g=(g_i|i\in V)$. It is convenient to introduce
a basis in $l_2(\7Z_d^V)$, called the computational basis:
$\{|\1g\rangle|\1g\in G^V\}$ where
$|\1g\rangle$ is the indicator function of the point $\1g$.

A weighted graph is given by the symmetric matrix
$\Gamma=(\Gamma(i,j)|i,j\in V)$, where $\Gamma(i,j)$ is the
integral number assigned to the edge $\{i,j\}$ being
zero if there is no edge between $i$ and $j$.
The cluster state, corresponding to $\Gamma$, is represented by a
normalized vector $\Psi^\Gamma$ in $l_2(\7Z_d^V)$. It is
given by assigning the value
\begin{eqnarray}\label{clusterI}
\Psi_\Gamma(\1g)
:=d^{-\frac{1}{2}|V|} \ \prod_{\{i,j\}\subset V}\chi(g_i|g_j)^{\Gamma(i,j)}
\end{eqnarray}
to a group element $\1g=(g_i|i\in V)$ in $\7Z_d^V$.
Here we have introduced for two group elements $g,h\in \7Z_d$ the
phase
\begin{equation}
\chi(g|h)=\exp\biggl(\frac{2\pi\8i}{d} \ g h \ \biggr) \ \ .
\end{equation}

A logical network for preparing a state, represented by a vector
$\Psi$ in $l_2(\7Z_d^V)$, is given a sequence of "quantum gates",
applied successively to a "ground" state
$|\10\rangle$ ($\10$ is the zero in $\7Z_d^V$), such that the resulting
vector is $\Psi$.
Quantum gates are elementary unitary
operations, each of them acting only on a single or two quantum
digits. In some cases operations on three quantum digits are also considered
as elementary gates, e.g. the Toffoli gate.

As it turns out, the following elementary gate operations are needed
for building a logical network for a cluster state:
\begin{itemize}
\item
The {\em local Fourier transform} $F_i$ operating on the
quantum digit at $i$
\begin{equation}
F_i|\1g\rangle:=\frac{1}{\sqrt{d}}\sum_{h\in\7Z_d} \
\chi(g_i|h) \ |h,\1g^i\rangle\ \ .
\end{equation}
with $\1g=(g_i|i\in V)$ and $\1g^i:=(g_i|i\in V\setminus \{i\})$
and its inverse $F_i^*$.
\item
The {\em $n$-controlled shift gate} $\1c(i,j)^n$ with control digit at position $i$
and target digit
at position $j$ is given by
\begin{equation}
\1c(i,j)^n|\1g\rangle:=|g_j+ng_i,\1g^j\rangle
\end{equation}
and its inverse $\1c(i,j)^{-n}$.
\end{itemize}

In graphical representation of
a logical network the local Fourier transform as well as
the $n$-controlled shift operation is symbolized as shown in Figure 1.
\begin{figure}[h]
\begin{center}
\epsfysize=3cm\epsffile{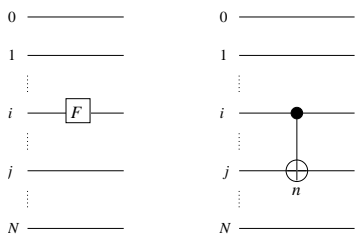}
\end{center}
\caption{From left to right:
The local Fourier transform at position $i$,
the $n$-controlled shift operation.}
\end{figure}
\begin{figure}[h]
\begin{center}
\epsfysize=3cm\epsffile{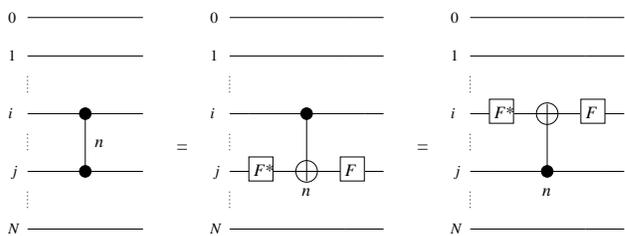}
\end{center}
\caption{The symbol for the $n$-controlled phase gate at positions $i,j$.}
\end{figure}

A procedure for preparing a cluster state (\ref{clusterI})
can directly be expressed in terms of local Fourier transforms and
{\em controlled phase gates} whose graphical symbols are given by Figure 2.
Each $n$-controlled phase gate
is a composition of two local Fourier transforms and one
$n$-controlled shift gate:
Acting on the positions $i,j$, this gate is given by
\begin{equation}\label{contphase}
\1u(i,j)^n:=F_j\1c(i,j)F_j^*=F_i\1c(j,i)^nF_i^* \ \ .
\end{equation}

According to the definition of the cluster state (\ref{clusterI}),
the vector $\Psi_\Gamma$ can be obtained by first applying
for each vertex a local Fourier transform and second operating with
a controlled phase gate on the
quantum digits corresponding to a pair of vertices which are
connected by an edge.
Introducing the {\em cluster state creation operator}
\begin{equation}\label{clusterIII}
\1u_\Gamma=\prod_{\{i,j\}\subset V}\1u(i,j)^{\Gamma(i,j)}  \
\prod_{j\in V} F_j
\end{equation}
the cluster state wave function is:
\begin{equation}\label{clusterII}
\Psi_\Gamma=\1u_{\Gamma}|\10\rangle \ \ .
\end{equation}

\begin{figure}[h]
\begin{center}
\epsfysize=3cm\epsffile{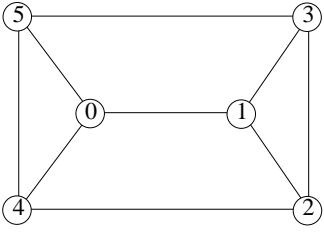}
\end{center}
\caption{The graph for the cluster state.}
\end{figure}

Considering the graph in Figure 3, the corresponding logical
network expressed in terms of controlled phase gates and local
Fourier transforms is given by Figure 4 below.

\begin{figure}[h]
\begin{center}
\epsfysize=3cm\epsffile{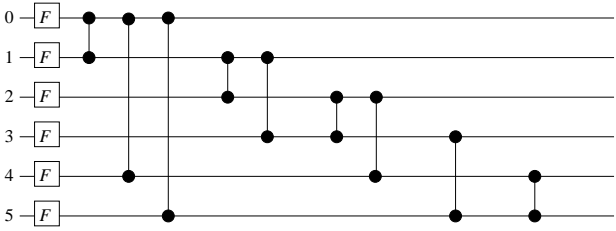}
\end{center}
\caption{Network in terms of controlled phase gates and
local Fourier transforms.}
\end{figure}

In order to obtain a logical network in terms of
controlled shift gates and local Fourier transforms, one just
have to substitute the controlled phase gates $\1u(i,j)$ in
Equation (\ref{clusterIII}) by
$F_j\1c(i,j)F_j^*$. This procedure is represented by Figure 5
for the cluster state corresponding to the graph in Figure 3.
As a consequence, we obtain a network implementation which uses a
local Fourier
transform for each vertex and a controlled shift gate for each
edge.

\begin{figure}[h]
\begin{center}
\epsfysize=7cm\epsffile{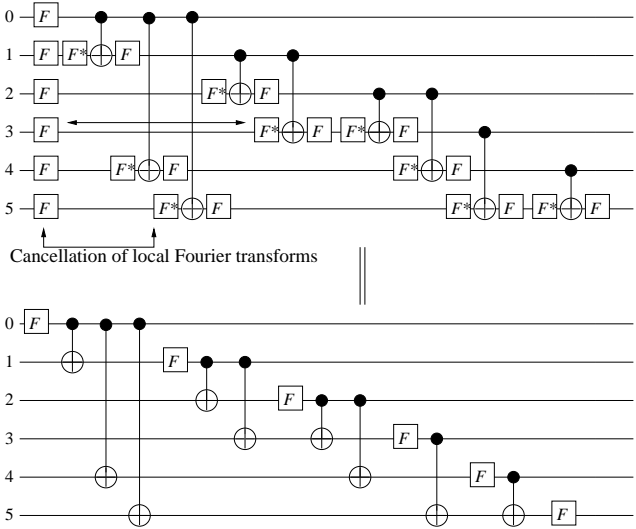}
\end{center}
\caption{After substituting the controlled phase gates in Figure 4
by controlled shift gates and local Fourier transforms, one
obtains the network implementation for the cluster state in terms
of controlled shift gates and local Fourier transforms.}
\end{figure}

\begin{Pro}\label{P1}
For a weighted graph $\Gamma$ with $v$ vertices and $l$ edges
the cluster state creation operator $\1u_\Gamma$ can be
decomposed into a product of $v+l$ unitary operations which are
local Fourier transforms or controlled shift gates.
\end{Pro}

We postpone the proof of the proposition to Appendix \ref{AI} and
describe an algorithm which associates to a weighted graph $\Gamma$ a
logical network expressing $\1u_\Gamma$
in terms of controlled shift gates and local Fourier transforms.
\begin{Alg}\label{alg1}{\em
Choose an enumeration of the vertices by
$V=\{0,\cdots,N\}$ and perform step 1 for the vertex $j=0$:
\begin{enumerate}
\item[1:]
Apply the Fourier transform $F_j$ and proceed with step 2
for the vertex $k=j+1$.
\item[2:]
Apply the controlled shift gate $\1c(j,k)^{\Gamma(j,k)}$.
\item[3:]
Repeat step 2 for the vertex $k'=k+1$ until $k'=N$.
\item[4:]
Repeat steps 1-3 by starting step 1 for the vertex $j'=j+1$ until
$j'=N$.
\end{enumerate}}
\end{Alg}
It will directly follow from the proof of Proposition \ref{P1}
that the Algorithm \ref{alg1} yields the correct decomposition
of $\1u_\Gamma$. In particular,
by applying it to the graph in Figure 3, one
obtains the logical network implementation presented in Figure 5.

\section{From cluster states to graph codes}
\label{SII}
In this section  we tackle the question, how graph
codes can be realized by preparing a cluster state
(a globally parallel operation) and local measurement operations.

To start with, we briefly describe here the concept of graph
codes, introduced in \cite{SchlWer00}. Consider a graph $\Gamma$
with vertices $V$ and choose
a subset of "input" vertices $X\subset V$ which reasonably
contains less elements than the set of "output" vertices
$Y=V\setminus X$. The quantum code, associated with this choice of
input and output vertices, is given by a linear map $\1v_\Gamma$
mapping the Hilbert space for the input register $l_2(\7Z_d^X)$
into the Hilbert space for the output register $l_2(\7Z_d^Y)$ as
follows: A basis vector $|\1h\rangle\in l_2(\7Z_d^X)$ with
$\1h\in\7Z_d^X$ is mapped to
\begin{equation}\label{graphcode}
\1v_\Gamma|\1h\rangle=d^{\frac{1}{2}|X|}\sum_{\1g\in
G^Y}\Psi_\Gamma(\1h,\1g) \ |\1g\rangle
\end{equation}
where $\Psi_\Gamma\in l_2(G^V)$ is the wave function for the
cluster state, associated with $\Gamma$ (\ref{clusterI}).
The range of $\1v_\Gamma$ is then a candidate for a protected
subspace corresponding to a quantum error correcting code.
In particular, if $\Gamma$ is a weighted graph, associated with a
quantum  error correcting code (See \cite{SchlWer00}),
then $\1v_\Gamma$ is automatically an isometry, i.e.
$\1v_\Gamma^*\1v_\Gamma=1$.

Concerning the Heisenberg picture, the isometry
$\1v_\Gamma$ implements a "coding channel" which is
the completely positive unital map,
assigning to an operator (observable of the output system)
$a$ on $l_2(\7Z_d^Y)$ to an observable of the input system:
\begin{equation}
a\mapsto\1v_\Gamma^* a \1v_\Gamma \ \ .
\end{equation}
For later purpose, it is convenient to introduce for a finite set
$V$ the algebra $\6A(V)$ of all bounded operators on
$l_2(\7Z_d^V)$. For a subset $K\subset V$ the algebra
$\6A(K)$ can be identified with the subalgebra of $\6A(V)$ which
consists of those operators acting nontrivially only on the
tensor factors corresponding to the subset $K$.
Moreover, we denote by $\6C(V)$ the abelian algebra of function
on $\7Z_d^V$, which can be identified with the algebra of
multiplication operators in $\6A(V)$.

The cluster state wave function occurs explicitly in the
formula for the graph code (\ref{graphcode}).
This suggests the following scheme for implementing the code by
preparing a cluster state and doing local measurement
operations:
\begin{enumerate}
\item[1:]
Prepare the input digits in the state which one wish to
protect against errors.
\item[2:] Prepare the output digits in the ground state
corresponding to the vector $|\10_Y\rangle$ (See (\ref{op1})
below).
\item[3:]
Apply to each input digit an inverse Fourier transform.
By acting on all quantum digits, apply
the cluster state creation operator $\1u_\Gamma$.
Operate with a Fourier transform on each input
(See (\ref{disdyn})).
\item[4:]
Perform a local measurement on each input
digit in the computational
basis (See (\ref{op3})).
\end{enumerate}

Concerning the scheme above, the coding procedure depends on the
outcome of the measurement of the input digits. Hence, besides the
quantum output register we obtain an additional classical output.
This can be modeled (in the Heisenberg picture)
by the unital completely positive
map $C_\Gamma$ which maps
the operator $a\otimes f$ in $\6A(Y)\otimes\6C(X)$ to
\begin{eqnarray}\label{op3}
C_\Gamma(a\otimes f):=d^{-|X|}\sum_{\1h\in\7Z_d^X}
\hat \1u(\1h)^*\1v_\Gamma^*
a\1v_\Gamma\hat \1u(\1h) \ f(\1h)
\end{eqnarray}
where $\1h\mapsto\hat\1u(\1h)$ is
the representation of $\7Z_d^X$ by multiplier operator
as given in Appendix \ref{AII}.

The following operations, given in terms of completely
positive maps are building blocks for performing $C_\Gamma$:
\begin{itemize}
\item
An automorphism $\alpha_\Gamma$
describes a discrete dynamics of the system
consisting of the input and output digits. It acts on the
corresponding observable algebra by
mapping an operator $b$ on
$l_2(\7Z^V_d)$ to
\begin{equation}\label{disdyn}
\alpha_\Gamma(b):=F_X\1u_\Gamma^*F_X^* \ b  \ F_X\1u_\Gamma F_X^* \ \ .
\end{equation}
Here $\1u_\Gamma$ is the cluster state
creation operator
(\ref{clusterIII}).
\item
The ground state preparation of the output digits $P_Y$ is
a completely positive unital map, sending an operator $b$, acting on
all quantum digits, to the operator
\begin{equation}\label{op1}
P_Y(b)=\1w_Y^*b\1w_Y
\end{equation}
acting on the inputs.
We introduce the isometry
$\1w_K$ by assigning to $\Psi\in l_2(\7Z_d^{V\setminus K})$ the vector
\begin{equation}\label{op2}
\1w_K\Psi:=|\10_K\rangle\otimes \Psi
\end{equation}
where $\10_K$
is the zero in $\7Z_d^K$.
\item
A measurement of the input digits in the computational basis,
corresponds to the unital
completely positive map $M_X$ mapping the operator $a\otimes f$ in
$\6A(Y)\otimes \6C(X)$ to
\begin{equation}\label{op3}
M_X(a\otimes f):=\sum_{\1h\in\7Z_d^X}\1u(\1h)\1w_X
a\1w_X^*\1u(\1h)^* \
f(\1h) \ \ .
\end{equation}
\end{itemize}

The following statement, which
we prove in Appendix \ref{AII}, justifies the suggested coding
scheme, described above:
\begin{Pro}\label{P2}
Let $\Gamma$ be a weighted graph, associated with a quantum error
correcting code.
By adopting the notations, given above, the isometry $\1v_\Gamma$
fulfills the identity
\begin{equation}\label{P2-1}
\1v_\Gamma=d^{\frac{|X|}{2}} \1w_X^*F_X\1u_\Gamma F_X^*\1w_Y \ \ .
\end{equation}
In particular, the coding operation $C_\Gamma$ satisfies
\begin{equation}\label{P2-2}
C_\Gamma= P_Y\circ\alpha_\Gamma\circ M_X \ \ .
\end{equation}
\end{Pro}

\section{Logical networks for graph codes I}
\label{SIII}
For getting the logical network, which performs
step 3 of the previous section,
we have to decompose the operator $F_X\1u_\Gamma F_X^*$ into a product
of controlled shift gates and local Fourier
transforms, which is a straight forward task since we know already
how to decompose the cluster state creation operator $\1u_\Gamma$.
As a result, we get with help of Proposition \ref{P1}:
\begin{Pro}\label{P3}
Let $\Gamma$ be a weighted graph,
associated with a quantum  error correcting code,
having $v$ vertices, $l$ edges and
no edges between input vertices. Then the
operator $F_X\1u_\Gamma F_X^*$ is a product of
$v+l$ elementary gates being either
controlled shift gates or local Fourier transforms.
\end{Pro}

As it is already mentioned in \cite{SchlWer00}, links between
input vertices can always be removed without affecting the error
correcting capabilities of the corresponding code. Therefore, the
assumptions in Proposition \ref{P3} (the proof can be
found in Appendix \ref{AIII}) can be made without loss of generality.

Analogously to the Algorithm \ref{alg1}, we also obtain an
algorithm for decomposing  the
operator $F_X\1u_\Gamma F_X^*$:

\begin{Alg}\label{alg2}{\em
Choose an enumeration of the input vertices by
$X=\{0,\cdots,k\}$ and the output vertices $Y=\{k+1,\cdots,v\}$.
Perform step 1 for the input vertices $j=0$ and $i=1$:
\begin{enumerate}
\item[1:]
Apply the controlled shift gate $\1c(j,i)^{\Gamma(j,i)}$.
\item[2:]
Repeat step 1 for the vertices $j$ and $i'=i+1$ until $i'=k$.
\item[3:]
Apply the Fourier transform $F_j$ and proceed with step 1
for the input vertices $j'=j+1$ and $i'=j+2$ if $i'\leq k$.
Otherwise, proceed with step 4.
\item[4:]
Apply the Algorithm \ref{alg1}, starting with the output vertex
$k+1$.
\end{enumerate}}
\end{Alg}

The graph in Figure 6 yields an example for a quantum error detecting
code, encoding two quantum digits $\{0,1\}$,
into four $\{2,\cdots,5\}$ and detecting all errors
which affect one quantum digit.
\begin{figure}[h]
\begin{center}
\epsfysize=3cm\epsffile{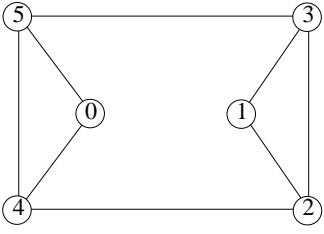}
\end{center}
\caption{The graph for a quantum error detecting code of length 4
encoding two quantum digits (inputs $\{0,1\}$) and detecting one error.}
\end{figure}
Performing the steps 1-3 of the Algorithm \ref{alg2},
for the graph in Figure 6, one obtains the part "Steps 1-3" of the logical
network depicted in Figure 7. The part "Step 4" is
obtained from an application of the Algorithm \ref{alg1} by
starting with vertex $\{2\}$. This is nothing else but the logical
network implementation for the
cluster state associated with
the subgraph
which is obtained by removing the input vertices $\{0,1\}$.

\begin{figure}[h]
\begin{center}
\epsfysize=3.5cm\epsffile{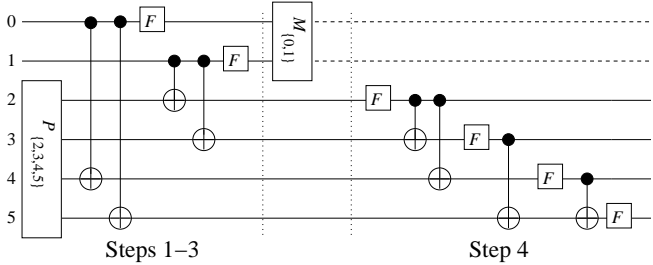}
\end{center}
\caption{Logical network for implementing the quantum error correcting code
corresponding to the graph of Figure 6 with inputs $\{0,1\}$.
The horizontal dashed lines correspond to classical wires.}
\end{figure}

\section{Logical networks for graph codes II}
\label{SIV}
In the previous section, we have given an algorithm which
associates a logical network to a weighted graph implementing the
corresponding quantum error correcting code.
Here the network operates on the input digits as well as on the output
digits. After applying the network, a local measurement of the inputs
is performed which completes the coding procedure.

In this section, we discuss the construction of local networks
for graph codes, which only operate on the output digits an which
do not require a measurement procedure after applying the network.

We give here a practicable solution for the case that
there is one input vertex. Furthermore we require that there
exists an output vertex being connected with the input by
an edge with weight $1$.

\begin{Pro}\label{P4}
Let $\Gamma$ be a weighted graph with $l$ edges, input vertex $\{0\}$ and
output vertices $\{1,\cdots,n\}$. If $\Gamma(0,1)=1$,
then there exists a unitary
operator $\1z_\Gamma$, acting on the output digits,
such that
\begin{equation}\label{p4-1}
\1v_\Gamma=\1z_\Gamma\1w_{\{2,\cdots,n\}} \ \ .
\end{equation}
Furthermore, $\1z_\Gamma$ can be decomposed into a product of
$l+n-1$ elementary gate operations which are either controlled
shifts or local Fourier transforms.
\end{Pro}

We prove the proposition in Appendix \ref{AIV}. The coding
operation is performed by the following two steps:

\begin{itemize}
\item
Prepare the output digit $\{1\}$ in the state one wishes to protect.
The remaining output digits are prepared in the state,
corresponding to the vector $|\10_{\{2,\cdots,n\}}\rangle$.
\item
Apply the logical network which implements the unitary operator
$\1z_\Gamma$.
\end{itemize}

The Algorithm \ref{alg3}, given below, associates to each
graph, which satisfies the assumptions of Proposition \ref{P4},
a decomposition of $\1z_\Gamma$ into elementary gates.

\begin{Alg}\label{alg3}{\em
Perform step 1 for the output vertex $i=2$:
\begin{enumerate}
\item[1:]
Apply the controlled shift gate $\1c(1,i)^{\Gamma(0,i)}$.
\item[2:]
Repeat step 1 for the vertex $i'=i+1$ until $i'=n$.
\item[3:]
Apply the Algorithm \ref{alg1}, starting with the output vertex
$j=1$.
\end{enumerate}}
\end{Alg}

We illustrate the statement of Proposition \ref{P4} by considering
the quantum error correcting code which corresponds to the graph
in Figure 3 with input vertex $\{0\}$.

\begin{figure}[h]
\begin{center}
\epsfxsize=8.5cm\epsffile{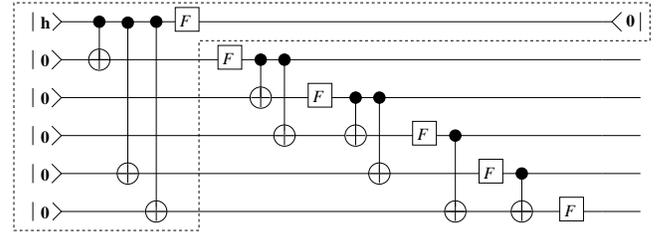}
\end{center}
\caption{Network representation of the
operator $d^{-1/2}\1v_\Gamma$ for
the graph in Figure 3 with input vertex $\{0\}$.}
\end{figure}

By Proposition \ref{P2}, the network in Figure 8 represents
the coding operation in the following way:
Each output digit $y=1,\cdots,5$ is prepared in the state $|0\rangle$, the input
digit $x=0$ is prepared in the state $|h\rangle$. After applying the
logical network, which implements $F_0\1u_\Gamma F_0^*$,
a selective measurement is performed on the input by collecting
those measurement outcomes for which each input digit is in the state
$|0\rangle$.
This selection procedure yields a factor $d^{-1/2}$ and
Figure 8 represents the operator $d^{-1/2}\1v_\Gamma$.

\begin{figure}[h]
\begin{center}
\epsfxsize=8.5cm\epsffile{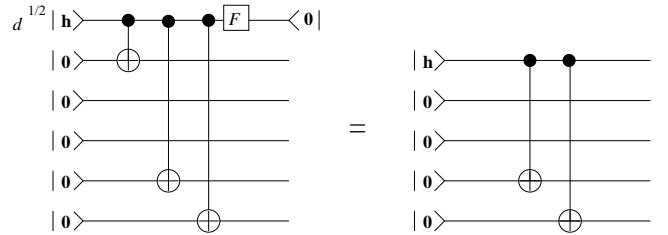}
\caption{Useful identity.}
\end{center}
\end{figure}

Now we make use of the identity represented in Figure 9. If we
replace in Figure 8 the part within the dashed frame by the
network on the right hand side of Figure 9, we obtain the
network depicted in Figure 10 below.
Due to the identity in Figure 9, we gain a factor $d^{1/2}$ and
Figure 10 represents the full isometry $\1v_\Gamma$ operating on the
basis vector $|h\rangle$, $h\in\7Z_d$.

\begin{figure}[h]
\begin{center}
\epsfxsize=8.5cm\epsffile{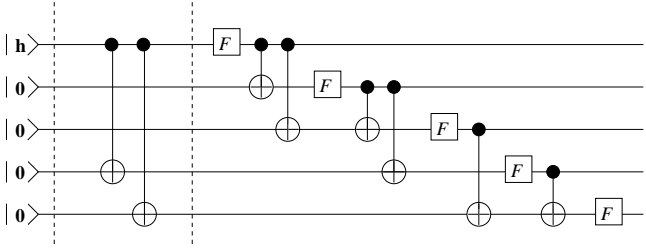}
\end{center}
\caption{Logical network, implementing the code, corresponding to the graph
in Figure 3 with input $\{0\}$.}
\end{figure}

By applying the Algorithm \ref{alg3} directly to the graph
in Figure 3 (with input vertex $\{0\}$) one indeed obtains the network in
Figure 10.

\section{Conclusion and outlook}
The ability of preparing
cluster states yields a resource for performing quantum
algorithms \cite{BrieRau00,BrieRau01}, on one hand, and it provides a
natural mechanism for protecting quantum information against
errors, on the other hand. In the present paper we have
developed the following features:

\begin{itemize}
\item
Logical networks for preparing general cluster states can be
derived from their
defining graphs in a systematic manner. The number of elementary
operations which is used by the network is the number of vertices
plus the number of edges of the graph.
\item
Quantum error correcting codes can be realized by
applying preparation procedure for a cluster state to a
suitably prepared input state
followed by a local measurement operation on
the input digits.
These coding schemes operates on the input and output digits and
they can be expressed in terms of logical networks
which uses the same amount of elementary gates as the
preparation procedure of the corresponding cluster state.
\end{itemize}

In order to save resources, one wishes to construct logical
networks, implementing quantum error correcting codes, by
operating only on a number of digits corresponding to the length
of the code (number of outputs). In fact we have given an
algorithm which handle the following situation:

\begin{itemize}
\item
Provided there is one input vertex, a logical network can be associated
to a given graph which implements the corresponding graph code.
This network operates directly on the output digits and its
number of elementary gate operations is the number of outputs plus
the number of edges minus one.
\end{itemize}

Concerning future investigations, it would be desirable to develop
similar network representations also for the decoding operations.
Here, one possible strategy is to look for a decoding procedure
which starts from preparing a cluster state, and then performing
local measurements on a suitable set of digits.
This problem can be tackled by searching for
a graph representation for decoding channels similar to those for
the coding operations.

\subsubsection*{{\it Acknowledgment:}}
I am grateful to R.F.Werner for supporting this
investigation with many ideas. I would also like to acknowledge
interesting and helpful discussions with
M.Grassl, H.Briegel and R.Raussendorf.
Funding by the European Union project
EQUIP (contract IST-1999-11053) is
gratefully acknowledged. This research project is also supported by
the Deutsche Forschungsgemeinschaft (DFG-Schwerpunkt
"Quanteninformationsverarbeitung").

\begin{appendix}
\section{Proof of Proposition \ref{P1}}
\label{AI}
Let $\Gamma$ be a weighted graph with vertices $V=\{1,\cdots,v\}$.
By introducing the operators
\begin{eqnarray}\label{cphaseblock}
\1u_\Gamma^{(j)}=\prod_{k=j+1}^{v}\1u(j,k)^{\Gamma(j,k)}
\end{eqnarray}
the cluster state creation operator (\ref{clusterIII}) can be
written as
\begin{equation}\label{decomp}
\1u_\Gamma=\1u_\Gamma^{(v-1)}\1u_\Gamma^{(v-2)}\cdots\1u_\Gamma^{(1)}
F_{\{1,\cdots,v\}} \ \ .
\end{equation}
We introduce for each vertex $j$ the block of controlled shift operations
\begin{eqnarray}\label{cnotblock}
\1c^\Gamma_j:=\prod_{i=j+1}^v\1c(j,i)^{\Gamma(j,i)} \ \ .
\end{eqnarray}
Note that $\1c(i,j)$ commutes with $\1c(i,k)$ which implies that
the definition of
$\1c^\Gamma_j$ is independent of the order of factors on the
right hand side of (\ref{cnotblock}).

The local Fourier transform $F_i$ commutes with the controlled
phase operation $\1u(j,k)$ (as well as the controlled shift $\1c(j,k)$)
for $i\not=j,k$. Thus we obtain from the definition of the
controlled phase gate (\ref{contphase})
and from (\ref{cphaseblock}):
\begin{eqnarray}\label{block}
\1u_\Gamma^{(j)}&=&F_v\1c(j,v)^{\Gamma(j,v)}F_{v}^*
\\ \nonumber
&&\times \ \1u(j,v-1)^{\Gamma(j,v-1)}\cdots\1u(j,j+1)^{\Gamma(j,j+1)}
\\ \nonumber
&=&
F_{v}\1c(j,v)^{\Gamma(j,v)}
F_{v-1}\1c(j,v-1)^{\Gamma(j,v-1)}F_{v-1}^*
\\ \nonumber
&&\times \
\1u(j,v-2)^{\Gamma(j,v-2)}\cdots\1u(j,j+1)^{\Gamma(j,j+1)}F_{v}^*
\\ \nonumber
&=&
F_{v-1}F_{v} \ \1c(j,v)^{\Gamma(j,v)}
\1c(j,v-1)^{\Gamma(j,v-1)}
\\ \nonumber
&&\times \
\1u(j,v-2)^{\Gamma(j,v-2)}\cdots\1u(j,j+1)^{\Gamma(j,j+1)} \ F_{v}^*F_{v-1}^*
\\ \nonumber
&=&
F_{\{j+1,\cdots,v\}} \ \1c^\Gamma_{j} \
F_{\{j+1,\cdots,v\}}^*
\end{eqnarray}
Inserting the identity (\ref{block}) into (\ref{decomp}) yields
\begin{eqnarray}\label{finaldecomp2}
\1u_\Gamma
&=&F_v \ \1c^\Gamma_{v-1}F_{v-1}\cdots
\1c_2^\Gamma F_2  \ \1c_1^\Gamma F_{\{2,\cdots,v\}}^*
F_{\{1,\cdots,v\}}
\\ \nonumber
&=&F_v \ \1c^\Gamma_{v-1}F_{v-1}\cdots
\1c_2^\Gamma F_2  \ \1c_1^\Gamma F_1 \ \ .
\end{eqnarray}
Looking at (\ref{cnotblock}) each block $\1c_{j}^\Gamma$ contains
$\sum_{k=j+1}^n 1^{\Gamma(j,k)}$ controlled shift operations. Therefore,
the total number of controlled shift operations
in the last line of
(\ref{finaldecomp}) is just the number of edges
$l$ of the graph. In addition to that,
for each vertex there is a elementary Fourier
transform and $\1u_\Gamma$ can
indeed be decomposed into a product of $v+l$
elementary gate operations being either local Fourier transforms or
controlled shift operations.
$\Box$
\section{Proof of Proposition \ref{P2}}
\label{AII}

\subsection{Shift and multiplier}
For a subset $K\subset V$, the group $\7Z_d^K$ can naturally be
identified with a subgroup in $\7Z_d^V$. The group $\7Z_d^K$ is
represented on $l_2(\7Z_d^K)$ by {\em shift operators}
according to
\begin{equation}
\1u(\1h)|\1h'\rangle=|\1h'+\1h\rangle
\end{equation}
with $\1h\in\7Z_d^K$ and $\1h'\in\7Z_d^V$.

By using the  Fourier transform $F_K=\prod_{k\in K}F_k$ on the digits
labeled by $K$, one obtains a further representation $\hat \1u$ of $\7Z_d^K$
by {\em multiplier operators}.
Namely, the multiplier $\hat\1u(\1h)$, $\1h\in\7Z_d^K$,
is related to the corresponding shift $\1u(\1h)$ according to
\begin{equation}
\hat\1u(\1h)|\1h'\rangle=F_K\1u(\1h)F_K^*|\1h'\rangle
=\chi(\1h|\1h')|\1h'\rangle \ \ .
\end{equation}
The multiplication phase is just given by
$\chi(\1h|\1h')=\prod_{k\in K}\chi(h_k|h_k')$.

\subsection{Proof of the proposition}
Let $|\1h\rangle$, $\1h\in\7Z_d^X$
be a vector of the computational basis of $l_2(\7Z^X_d)$.
The cluster state creation operator $\1u_\Gamma$ is the composition of the
Fourier transform on $F_V$ on the full register of all quantum digits
and the unitary multiplication operation $\Phi_\Gamma$, given by
\begin{equation}\label{PRP2-1}
\Phi_\Gamma|\1g\rangle:=d^{\frac{|V|}{2}}\Psi_\Gamma(\1g)|\1g\rangle
\end{equation}
where $\Psi_\Gamma$ is the cluster state wave function  (\ref{clusterI})
associated with the weighted graph $\Gamma$.
Furthermore, we have $\1w[\1h]=\1u(\1h)\1w_X$ for each
$\1h\in\7Z_d^X$.
Thus, the right hand side of (\ref{P2-1}) can be written as
\begin{equation}\label{PRP2-2}
\1w_X^*F_X\1u_\Gamma
F_X^*\1w_Y=\1w_X^* F_X\Phi_\Gamma F_Y\1w_Y \ \ .
\end{equation}
which implies for each $\1h\in\7Z_d^X$.
\begin{eqnarray}\label{PRP2-2-1}
\1w_X\1u(\1h)^*F_X\1u_\Gamma F_X^*\1w_Y
&=&\1w_X^*\1u(\1h)^*F_X\Phi_\Gamma F_Y\1w_Y
\\ \nonumber
&=&\1w_X^*F_X \hat\1u(\1h)^*\Phi_\Gamma F_Y\1w_Y
\\ \nonumber
&=&\1w_X^*F_X\Phi_\Gamma \hat\1u(\1h)^* F_Y\1w_Y
\\ \nonumber
&=&\1w_X^*F_X\Phi_\Gamma F_Y\1w_Y \hat\1u(\1h)^*
\end{eqnarray}
where we have used the fact that both $\hat\1u(\1h)$
and $\Phi_\Gamma$ are multiplication operators.
Furthermore, $\hat\1u(\1h)$ acts only on the input digits
and therefore it commutes with the operator $F_Y\1w_Y$, which only
affects on the output digits.

Now the Fourier transform $F_Y$ on the output digits maps the
vector $\1w_Y|\1h\rangle=|\1h,\10_Y\rangle$ to
\begin{eqnarray}\label{s2}
F_Y\1w_Y
|\1h\rangle=d^{-\frac{1}{2}|Y|}\sum_{\1g\in\7Z_d^Y}\ |\1h,\1g\rangle
\end{eqnarray}
and an application of the operator $\Phi_\Gamma$ to
(\ref{s2}) yields the expression
\begin{eqnarray}\label{s3}
\Phi_\Gamma F_Y\1w_Y|\1h\rangle=d^{\frac{|X|}{2}}
\sum_{\1g\in\7Z_d^Y}\Psi_\Gamma(\1h,\1g) |\1h,\1g\rangle \ \ .
\end{eqnarray}
Acting with the Fourier transform $F_X$ on (\ref{s3}) and
applying the co-isometry $\1w_X^*$ afterwards gives
\begin{eqnarray}\label{s4}
&&\hskip-15pt\1w_X^*F_X\Phi_\Gamma F_Y\1w_Y|\1h\rangle
\\ \nonumber
&=&\sum_{(\1h',\1g)\in\7Z_d^V} \ \chi(\1h|\1h')
\ \Psi_\Gamma(\1h,\1g) \1w_X^*|\1h',\1g\rangle
\\ \nonumber
&=&\sum_{(\1h',\1g)\in\7Z_d^V} \ \chi(\1h|\1h')
\ \Psi_\Gamma(\1h,\1g) \ \delta(\1h') \ |\1g\rangle
\\ \nonumber
&=&\sum_{\1g\in\7Z_d^Y} \ \Psi_\Gamma(\1h,\1g) \ |\1g\rangle
=d^{-\frac{|X|}{2}}\1v_\Gamma|\1h\rangle \ \ .
\end{eqnarray}
Here $\delta$ is the indicator function on $\7Z_d^X$
of the zero element $\10_X$. Note that
the co-isometry $\1w_X^*$ maps the vector $|\1h,\1g\rangle$ to
$\delta(\1h)|\1g\rangle$ for each $\1h\in\7Z_d^X$ and $\1g\in\7Z_d^Y$.

Finally, the identity (\ref{P2-2})  follows directly from
(\ref{P2-1}), (\ref{PRP2-2-1}), (\ref{s4}),
the definition of the discrete dynamics
$\alpha_\Gamma$ (\ref{disdyn}), the preparation
of the outputs $P_Y$ (\ref{op1}) and the measurement of the
inputs $M_X$ (\ref{op3}):
\begin{eqnarray}
&&\hskip-15pt P_Y\circ \alpha_\Gamma\circ M_X(a\otimes f)
\\\nonumber
&=&
d^{-|X|}\sum_{\1h\in\7Z_d^X} \hat\1u(\1h)^*\1v_\Gamma^*a\1v_\Gamma
\hat\1u(\1h) \ f(\1h)
\\\nonumber
&=&C_\Gamma(a\otimes f)
\end{eqnarray}
$\Box$
\section{Proof of Proposition \ref{P3}}
\label{AIII}
Let $\Gamma$ be a weighted graph with
input vertices $X=\{1,\cdots,k\}$
and output vertices $\{k+1,\cdots,v\}$.
By (\ref{finaldecomp}),
the cluster state creation operator $\1u_\Gamma$ can be written as
\begin{eqnarray}\label{finaldecomp}
\1u_\Gamma
=F_v \ \1c^\Gamma_{v-1}F_{v-1}\cdots
\1c_2^\Gamma F_2  \ \1c_1^\Gamma F_1
\end{eqnarray}
where the blocks of controlled shift operations $\1c_j^\Gamma$ are
given by (\ref{cnotblock}). Since we assume that there are no
edges between input vertices, for each input vertex $x\in X$,
the operator $\1c_x^\Gamma$ is of the form
\begin{equation}
\1c_x^\Gamma= \prod_{y\in Y}\1c(x,y)^{\Gamma(x,y)} \ \ .
\end{equation}
As a consequence $\1c_x^\Gamma$ commutes with all local Fourier transforms
$F_{x'}$ on those inputs $x'\in X$ for which $x'\not=x$.
Furthermore, the Fourier transform on the inputs $F_X$
commutes with
\begin{eqnarray}\label{outputdecomp}
F_v \ \1c^\Gamma_{v-1}F_{v-1}\cdots
\1c_{k+2}^\Gamma F_{k+2}  \ \1c_{k+1}^\Gamma F_{k+1} \ .
\end{eqnarray}
Hence we obtain
\begin{eqnarray}\label{dec}
F_X\1u_\Gamma F_X^*
&=&F_{\{1,\cdots,k\}}F_v\1c^\Gamma_{v-1}F_{v-1}\cdots
\1c_2^\Gamma F_2\1c_1^\Gamma F_1 F_{\{1,\cdots,k\}}^*
\\ \nonumber
&=&F_v \ \1c^\Gamma_{v-1}F_{v-1}\cdots
\1c_{k+2}^\Gamma F_{k+2}  \ \1c_{k+1}^\Gamma F_{k+1}
\\ \nonumber
&&\times \ F_{\{1,\cdots,k\}}\1c^\Gamma_{k}\cdots
\1c_2^\Gamma \1c_1^\Gamma
\\ \nonumber
&=&F_v \ \1c^\Gamma_{v-1}F_{v-1}\cdots
\1c_{k+2}^\Gamma F_{k+2}  \ \1c_{k+1}^\Gamma F_{k+1}
\\ \nonumber
&&\times
F_k\1c^\Gamma_{k}\cdots
F_2 \1c_2^\Gamma  \ F_1 \1c_1^\Gamma \ \ .
\end{eqnarray}
Now we see from (\ref{dec}) that the operator $F_X\1u_\Gamma F_X^*$
is a product of $v+l$ elementary gates, namely a
local Fourier transform for each
vertex, and a controlled shift gate for each edge.
$\Box$
\section{Proof of Proposition \ref{P4}}
\label{AIV}
Given a weighted graph $\Gamma$ with input vertex by $\{0\}$
and output vertices $\{1,\cdots,n\}$.
According to Proposition \ref{P2},
the isometry $\1v_\Gamma$ acts on a basis vector
$|h\rangle$, $h\in\7Z_d$, according to
\begin{equation}
\1v_\Gamma|h\rangle=d^{\frac{1}{2}}\1w_0^*F_0^*\1u_\Gamma F_0
|h,\10_{\{1,\cdots,n\}}\rangle \ \ .
\end{equation}
Making use of the identity (\ref{finaldecomp2}) we find
\begin{eqnarray}
\1v_\Gamma|h\rangle
=d^{\frac{1}{2}} F_n \ \1c^\Gamma_{n-1}F_{n-1}\cdots
\1c_1^\Gamma F_1  \ \1w_0^*F_0\1c_0^\Gamma|h,\10_{\{1,\cdots,n\}}\rangle \ \ .
\end{eqnarray}
Here we have used the fact that for any operator $a$, acting on
the outputs $\{1,\cdots,n\}$, we have
$\1w_0^*(\11\otimes a)=a\1w_0^*$.
Now we compute
\begin{eqnarray}
&&\hskip-20pt d^{\frac{1}{2}}\1w_0^*F_0\1c_0^\Gamma|h,\10_{\{1,\cdots,n\}}\rangle
\\   \nonumber
&=&
\1w_0^*F_0\prod_{y=k+1}^{k+n}\1c(0,y)^{\Gamma(0,y)}
|h,\10_{\{1,\cdots,n\}}\rangle
\\ \nonumber
&=&
\sum_{g\in\7Z_d}\chi(h|g) \ \delta(g)
|\Gamma(0,1)h,\cdots,\Gamma(0,n)h\rangle
\\ \nonumber
&=&
|\Gamma(0,1)h,\cdots,\Gamma(0,n)h\rangle
\end{eqnarray}
which implies by assuming $\Gamma(0,1)=1$:
\begin{eqnarray}
&&\hskip-20pt d^{\frac{1}{2}}\1w_0^*F_0\1c_0^\Gamma
|h,\10_{\{1,\cdots,n\}}\rangle
\\   \nonumber
&=&
\prod_{y=2}^n \1c(1,y)^{\Gamma(0,y)}|h,\10_{\{2,\cdots,n\}}\rangle
\\   \nonumber
&=& \1b_0^\Gamma \1w_{\{2,\cdots,n\}}|h\rangle \ \ .
\end{eqnarray}
Here we have introduced the operator
\begin{equation}
\1b_0^\Gamma:=\prod_{y=2}^n \1c(1,y)^{\Gamma(0,y)} \ \ .
\end{equation}
As a consequence, the operator
\begin{equation}
\1z_\Gamma:= F_n \ \1c^\Gamma_{n-1}F_{n-1}\cdots
\1c_1^\Gamma F_1 \ \1b_0^\Gamma
\end{equation}
fulfills (\ref{p4-1}). In particular, $\1z_\Gamma$
is a product of $l+n-1$ elementary gates.
$\Box$
\end{appendix}


\begin{thebibliography}{References}
\bibitem{l2}
For a general finite abelian group $A$, the
Hilbert space $l_2(A)$ consists of all complex-valued functions
on $A$. The scalar product is
\begin{equation}
\langle\Psi_0,\Psi_1\rangle=\sum_{a\in A}\overline{\Psi_0(a)}\Psi_1(a)\ \ .
\end{equation}
\end{thebibliography}
\end{document}